\documentclass[11pt,twoside,dvips]{article}
\usepackage{verbatim}
\usepackage{amssymb}
\usepackage{amsfonts}
\usepackage{amsmath}
\usepackage{pifont}
\usepackage[spanish,english]{babel}
\usepackage[pdftex]{graphicx}
\usepackage{verbatim}
\usepackage{epsfig}
\usepackage{bm}
\usepackage{longtable}
\usepackage{fancyhdr}
\begin{document}

\fancypagestyle{plain}{%
\fancyhf{}%
\fancyhead[LO, RE]{XXXVIII International Symposium on Physics in Collision, \\ Bogot\'a, Colombia, 11-15 september 2018}}

\fancyhead{}%
\fancyhead[LO, RE]{XXXVIII International Symposium on Physics in Collision, \\ Bogot\'a, Colombia, 11-15 september 2018}

\title{Testing the inert Zee model using lepton flavor observables }
\author{Robinson Longas$\thanks{%
e-mail: robinson.longas@udea.edu.co}$, 
\\ Instituto de F\'isica, Universidad de Antioquia, \\ Calle 70 No. 52-21, A.A. 1226, Medell\'in, Colombia}
\date{}
\maketitle

\begin{abstract}
The inert Zee model is an extension  of the Zee model for neutrino masses  to allow for a solution to the dark matter problem that involves two vector-like fields, a doublet and a singlet of ${\rm SU(2)_L}$, and two scalars, also a doublet and a singlet of ${\rm SU(2)_L}$, all of them being odd under an exact ${\rm Z_2}$ symmetry.
 The introduction of the ${\rm Z_2}$ guarantees one-loop neutrino masses, forbids tree-level Higgs-mediated flavor changing neutral currents and ensures the stability of the dark matter candidate.
 Due to the natural breaking of lepton numbers in the inert Zee model, we study the phenomenology of the processes leading to these kind of signals and establish which are the most promising experimental perspectives on that matter.  
\end{abstract}

\section{Introduction}
\label{sec:introduction}

Neutrino oscillations provide a clear evidence for lepton flavor violation (LFV) in the neutral sector, pointing out to physics beyond the Standard Model (SM). 
However, no evidence of lepton flavor violating processes in the charged sector has been found.
For instance, MEG collaboration has reported an upper limit on the decay branching ratio for $\mu\to e\gamma$ around $6\times10^{-13}$ \cite{Adam:2013mnn}, which will be improved soon by a factor of 10. Concerning the three-body decay $\mu \rightarrow 3e$, the negative searches for rare decays in the SINDRUM experiment lead to an upper limit for the branching ratio of 
around $10^{-12}$ \cite{Bellgardt:1987du}, whereas the Mu3e experiment collaboration  expects to reach the ultimate sensitivity to test such a decay in $10^{16}$ muon decays \cite{Blondel:2013ia}. In addition, the neutrinoless $\mu$-$e$ conversion in muonic atoms is also a promising way to search for charged LFV (CLFV) signals due to the significant increase of sensitivity expected for this class of experiments \cite{Carey:2008zz}.
Last but not least, the future plans regarding electron electric dipole moment (eEDM) \cite{Kawall:2011zz} are also in quest for New Physics signals since  the expected sensitivity for these facilities will improve by two orders of magnitude the current bound $|d_e|<8.7\times10^{-29} e\,\cdot\,$cm. 

On the other hand, despite the abundant evidence for the massiveness of neutrinos, the underlying mechanism behind it remains unknown, which is not a bizarre occurrence since the particle theory responsible for the dark matter (DM) of the Universe also resists to be experimentally elucidated. Hence, it would desirable that both phenomena may have a common origin. In this work we consider the inert Zee model (IZM) \cite{Longas:2015sxk}, a  DM realization of the Zee model for neutrino masses~\cite{Zee:1980ai}.
Neutrino masses are generated at one loop, while DM is addressed as in the inert doublet model (IDM) \cite{Barbieri:2006dq}. 
Once we will have determined the viable parameter space consistent with DM, neutrino oscillation observables, LFV processes and electroweak precision tests, we will establish the most relevant experimental perspectives regarding LFV searches.
Furthermore, since the Yukawa couplings that reproduce the neutrino oscillation data are complex, we will look into the regions in the parameter space where the prospects for the eEDM are
within the future experimental sensitivity.

\section{The model}
\label{sec:model}

We extend the Standard Model with the new particles shown in Table \ref{tab:content}, which consists of two vectorlike fermions, a ${\rm SU(2)_L}$-singlet $\epsilon$ and a ${\rm SU(2)_L}$-doublet $\Psi=(N, E)^{\text{T}}$, and two scalar multiplets, a ${\rm SU(2)_L}$-singlet $S^-$ and a ${\rm SU(2)_L}$-doublet $H_2=(H_2^+, H_2^0)^{\text{T}}$.
All of them are odd under the ${\rm Z_2}$ symmetry, which in turn allows us to avoid Higgs-mediated flavor changing neutral currents at tree-level, forbid tree-level contributions to the neutrino masses  and render the lightest ${\rm Z_2}$-odd particle stable \cite{Longas:2015sxk}. 
\begin{table}[h!]
\begin{center}
\begin{tabular}{|c|c|c|}
\hline
\hline\rule[0cm]{0cm}{.9em}
   & \text{Spin} & $ {\rm SU(3)_C \otimes SU(2)_L \otimes U(1)_Y \otimes Z_2}$ \\ 
\hline
\hline
$\epsilon$ & 1/2 & $({\bf 1},{\bf 1},-2,-1)$ \\\hline
$\Psi$ & 1/2 & $({\bf 1},{\bf 2},-1,-1)$ \\\hline
$H_2$ & 0 & $({\bf 1},{\bf 2},1,-1)$ \\\hline
$S$ & 0 & $({\bf 1},{\bf 1},-2,-1)$ \\\hline
\end{tabular}
\end{center}
\caption{The new particle content of the model with their transformation properties under the ${ \rm SU(3)_C \otimes  SU(2)_L \otimes U(1)_Y  \otimes Z_2}$ symmetry.}
\label{tab:content}
\end{table}

The most general ${\rm Z_2}$-invariant Lagrangian of the model is given by
\begin{align}
  \label{eq:Lmodel}
\mathcal{L}_{\text{IZM}} \supset \mathcal{L}_{\text{SM}}+\mathcal{L}_S+ \mathcal{L_Y},
\end{align}
where $\mathcal{L}_{\text{SM}}$ is the SM Lagrangian which includes the Higgs potential. $\mathcal{L}_S$ comprises, respectively, the mass, self-interacting and the interaction terms of the new scalars,
\begin{align}
   \mathcal{L}_S&=-\mu_2^2H_2^\dagger H_2-\frac{\lambda_2}{2} (H_2^\dagger H_2)^2 -\mu_S^2S^\dagger S-\lambda_S (S^\dagger S)^2 +
   \lambda_3 ( H_1^{\dagger}H_1 )( H_2^{\dagger}H_2 ) \nonumber\\ & + \lambda_4 ( H_1^{\dagger}H_2 )( H_2^{\dagger}H_1 )  +  \frac{\lambda_5}{2}\left[( H_1^{\dagger} H_2 )^2 + {\rm h.c.} \right] + \lambda_6 (S^\dagger S) (H_1^{\dagger}H_1) \nonumber\\
   & + \lambda_7 (S^\dagger S)(H_2^{\dagger}H_2) +  \mu  \epsilon_{ab}\left[H_1^a H_2^b S + {\rm h.c.} \right],
 \end{align} 
where $\epsilon_{ab}$ is the ${\rm SU(2)_L}$ antisymmetric tensor, {\small $H_2=(H_2^+, H_2^0)^{\text{T}}$} with {\small $H_2^0=(H^0 + i A^0)/\sqrt{2}$}. $H_2^0$ does not develop a vacuum expectation value in order to ensure the conservation of the ${\rm Z_2}$ symmetry. 
Finally, $\mathcal{L_Y}$ includes the new Yukawa interaction terms:
\begin{align}
   \label{eq:YukawaLagrangian}
   -\mathcal{L_Y}&=  \eta_i\bar{L}_{i}H_2\epsilon + \rho_i \bar{\Psi}H_2 e_{Ri} + y \bar{\Psi} H_1 \epsilon 
                   + f^*_i \overline{L^c_{i}} \Psi S^+  + {\rm h.c}.
 \end{align}
After electroweak symmetry breaking, the ${\rm Z_2}$-odd scalar spectrum consists of a CP-even state $H^0$, a CP-odd state $A^0$ and two charged states $\kappa_{1,2}$. On the other hand, the ${\rm Z_2}$-odd fermion spectrum involves two  charged fermions $\chi_{1,2}$ along with a neutral Dirac fermion $N$.

With respect to DM in the IZM, $H^0$ is the DM candidate as long as it remains as the lightest ${\rm Z_2}$-odd particle in the spectrum.
Hence, we expect the DM phenomenology to be similar to the one in the IDM in scenarios where the particles not belonging to the IDM 
do not participate in the DM annihilation processes \cite{Longas:2015sxk,Longas:2015cnr}.  
Accordingly, the viable DM mass range for this scenario is divided into two regimes \cite{Barbieri:2006dq}: the low mass regime, $m_{H^0} \simeq m_{h}/2$, and the high mass regime, $m_{H^0}\gtrsim 500$ GeV. 

The Majorana neutrino mass matrix in the mass eigenstates is given by 
\begin{align}\label{eq:neutrinomass}
  [M_{\nu}]_{ij} &= (\sin2\alpha \sin2\delta)/(64 \pi^2)(I_2 m_{\chi_2}-I_1 m_{\chi_1} ) [\eta_i f_j + \eta_j f_i ],
\end{align}
where $I_1$ and $I_2$ are loop functions \cite{Longas:2015sxk}. Using Eq.\eqref{eq:neutrinomass} and the diagonalization condition  
$U^{\text{T}}M_{\nu}U ={\rm diag}(m_1,m_2,m_3)$ with  $U=VP$ and  $P=\mbox{diag}(1,e^{i\phi/2},1)$, it is possible to express five of the six Yukawa couplings $\eta_i$ and $f_i$ in terms of the neutrino low energy observables,
\begin{align}\label{eq:yuks-f-eta}
  &\eta_i=|\eta_1|\frac{A_i}{\beta_{11}}, \hspace{1cm}f_i=\frac{1}{2\zeta}\frac{\beta_{ii}}{\eta_i},
\end{align}
where the coefficients, $\beta_{ij}$ and $A_{j}$ are given in \cite{Longas:2015sxk}.

\section{Charged lepton processes}
\label{sec:CLp}

In the IZM, CLFV processes such as $\ell_{i} \rightarrow \ell_{j}\gamma$, $\ell_{i} \rightarrow 3\ell_{j}$ and $\mu-e$ conversion in nuclei are generated at one-loop level involving the $\eta_{i}$, $f_{i}$ and $\rho_{i}$ Yukawa interactions in Eq.\eqref{eq:YukawaLagrangian}.
For $\ell_{i} \rightarrow \ell_{j}\gamma$, the branching ratios are given by
\begin{align}
\label{eq:bramuegamma}
 \mathcal{B}\left(\ell_{i} \rightarrow \ell_{j} \gamma \right)
 = \frac{3\alpha_{em}}{64\pi m_{\mu}^2G_F^2} \left( \left|\Sigma_L\right|^2 + \left|\Sigma_R\right|^2 \right) \mathcal{B}\left(\ell_{i} \rightarrow \ell_{j} \nu_i \bar{\nu_j} \right),
\end{align}
where $\alpha_{em}$ is the electromagnetic fine structure constant, $G_F$ is the Fermi constant and $\Sigma_L$, $\Sigma_R$ are given in \cite{Longas:2015sxk}. Concerning the $\ell_i \rightarrow \ell_j\bar{\ell_j}\ell_j$ processes, there are two class of diagrams: the $\gamma$- and $Z$- penguin diagrams and the box diagrams.  
Finally, the $\mu-e$ conversion diagrams are obtained when the pair of lepton lines attached to the photon and $Z$ boson in the penguin diagrams are replaced by a pair of light quark lines. There are no box diagrams since the ${\rm Z_2}$-odd particles do not couple to quarks at tree level, see ref. \cite{Gaviria:2018cwb} for details.

The Yukawa parameters in Eq. \eqref{eq:yuks-f-eta} required for  neutrino oscillation data are in general complex and constitute new sources of CP violation in the lepton sector. Thus the IZM has new contributions to the EDM of charged leptons at one-loop level given by 
\begin{equation} \label{eq:EDM}
 d_{\ell_i} = \frac{e}{2^6 \pi^2} s_{2\alpha} \operatorname{Im}(\rho_i \eta_i) \left[m_{\chi_2}\mathcal{I}_1( m^2_{\chi_2}, m_{H^0}^2, m_{A^0}^2 )-
 m_{\chi_1}\mathcal{I}_1( m^2_{\chi_1}, m_{H^0}^2, m_{A^0}^2 )\right],
\end{equation}
where the loop function is given in \cite{Gaviria:2018cwb}.

\section{Results and discussion}
\label{sec:numer-results-disc}

We have used {\tt SARAH} \cite{Staub:2013tta}, {\tt SPheno} \cite{Porod:2003um} and {\tt MicrOMEGAS} \cite{Belanger:2013oya}
to perform the numerical analysis. We consider both normal and inverted hierarchies for the neutrino mass spectrum and take the current best fit values reported in Ref.~\cite{deSalas:2017kay} and we have varied the free parameters as
\begin{align}
  \label{eq:scanLFVlow2}
  & 10^{-5} \leq | \eta_1 |,|\rho_1|,|\rho_2|,|\rho_3| \leq 3\;;\;\nonumber\\
  &0 \leq \text{Arg}(\eta_1), \text{Arg}(\rho_1), \text{Arg}(\rho_2), \text{Arg}(\rho_3)\leq2\pi\;;\nonumber\\
  & 100 \, {\rm GeV} \leq m_{A^0},\, m_{\kappa_1},m_{\chi_1}  \leq 500\, {\rm GeV}\;;m_{H^0}=60\,{\rm GeV}\; \;\nonumber\\
  &m_{\kappa_2} =  [m_{\kappa_1},600\,{\rm GeV}]\, 
    \;;\; m_{\chi_2} =  [m_{\chi_1},600\, {\rm GeV}]\;;\; \lambda_L\sim3\times10^{-4}\,.
\end{align}
The scan above satisfy the DM relic density~\cite{Ade:2015xua}, the $S$, $T$ and $U$ oblique parameters \cite{Baak:2014ora} and the LEP II constraints on the masses of the charged ${\rm Z_2}$-odd fermions and scalars \cite{Achard:2001qw}.

\begin{figure}[h!]
\centering
  \includegraphics[scale=0.45]{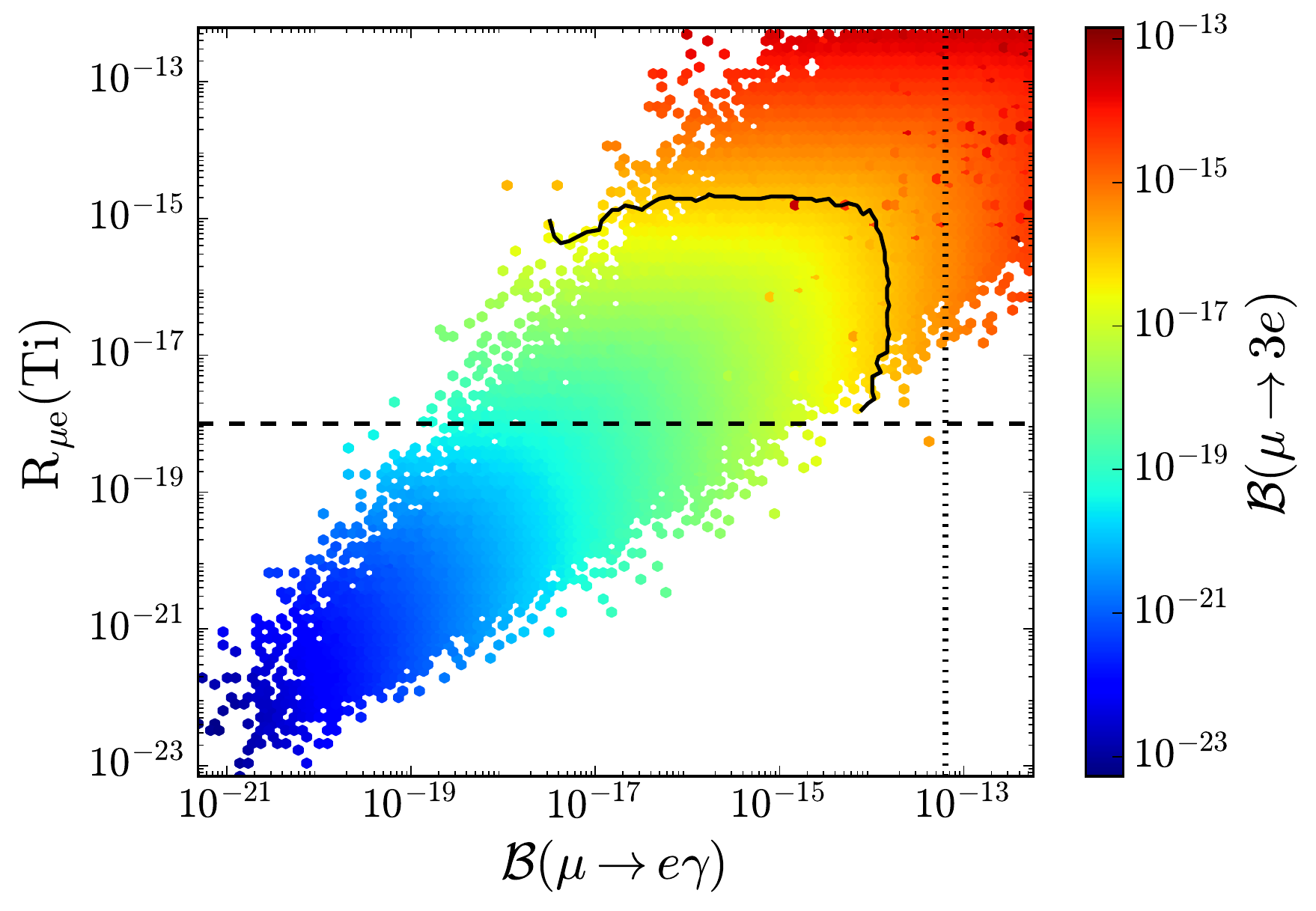}
  \includegraphics[scale=0.45]{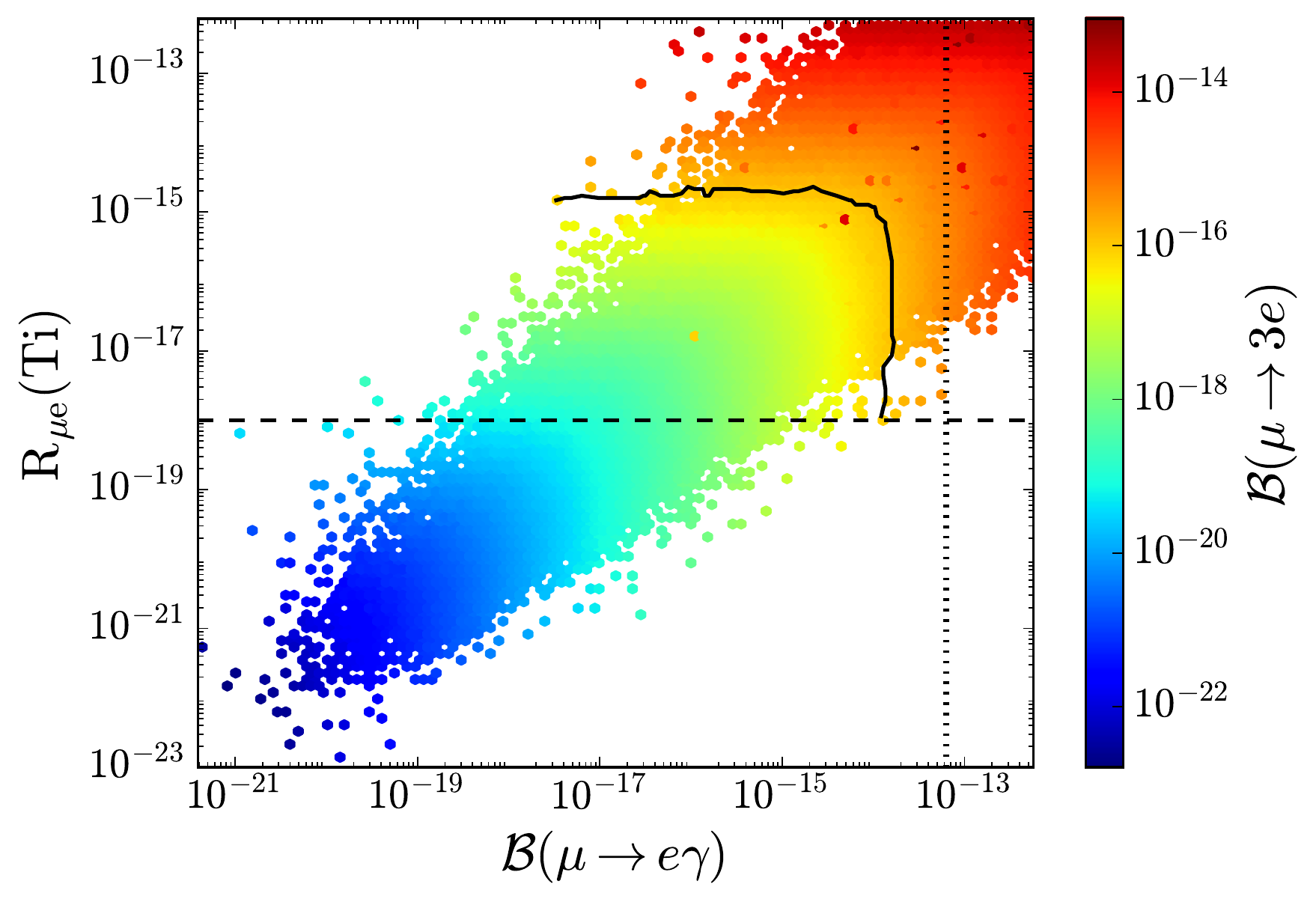}
  \caption{{\small The available parameter space of the IZM and the future exclusion zones coming from the most constraining CLFV searches. Left (right) panel is for NH (IH). The dotted, solid and dashed lines represent the expected sensitivity for the future searches regarding $\mathcal{B}(\mu\rightarrow e \gamma)$, $\mathcal{B}(\mu\rightarrow 3e)$ and $\mathcal{R}_{\mu e}$, respectively.}}
  \label{fig:LFV1}
\end{figure}

In Fig.~\ref{fig:LFV1} we show the correlation between the $\mathcal{B}(\mu\rightarrow e \gamma)$, $\mathcal{B}(\mu\rightarrow 3e)$ and $\mathcal{R}_{\mu e}$ observables, and the impact of the future searches associated to these observables over the parameter space considered. 
It follows that a large fraction of the current viable parameter space will be tested in the future experiments, with the $\mathcal{R}_{\mu e}$ observable being the most promising. 

\begin{figure}[h!]
  \begin{center}
    \includegraphics[scale=0.45]{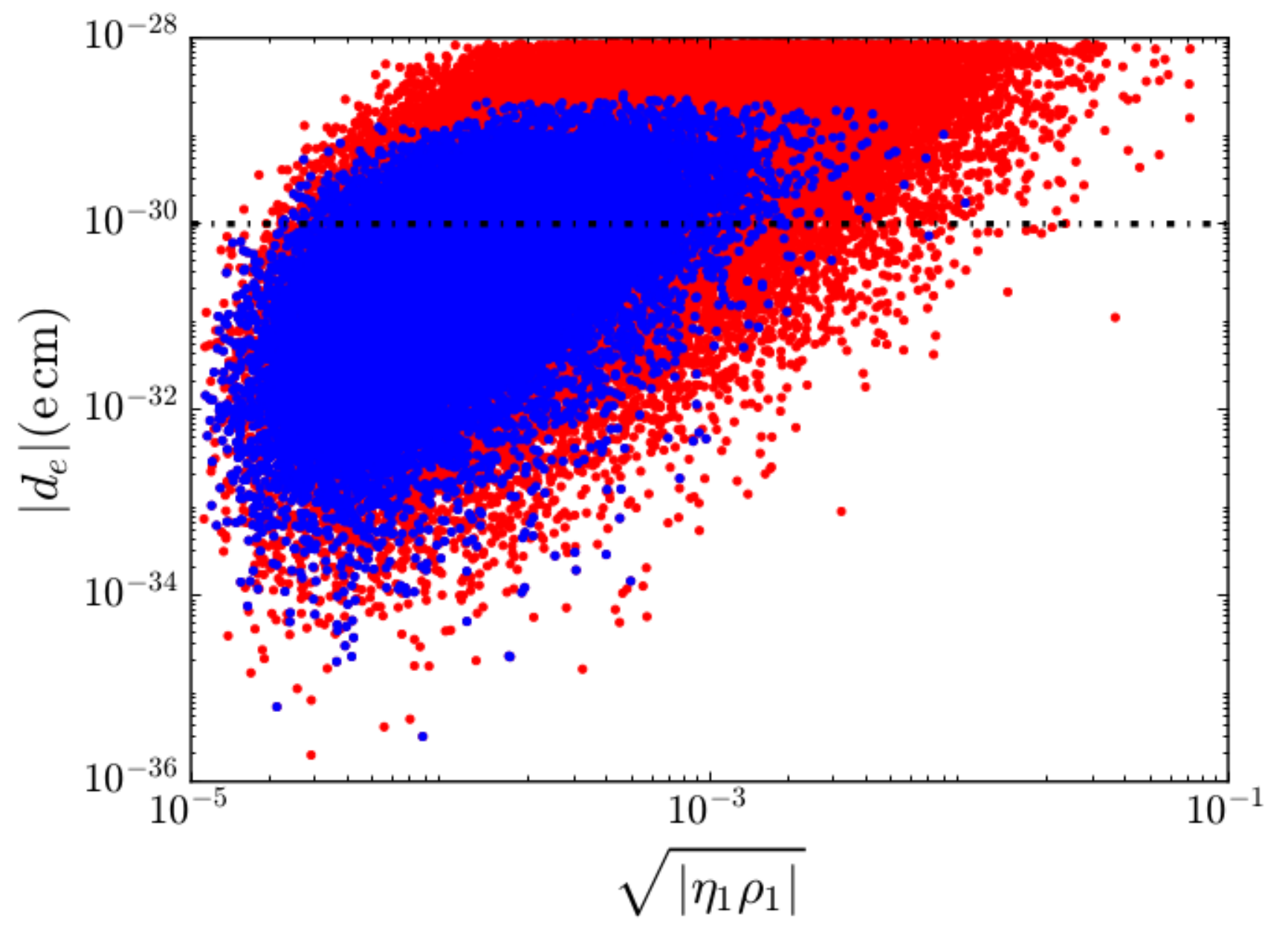}
    \includegraphics[scale=0.45]{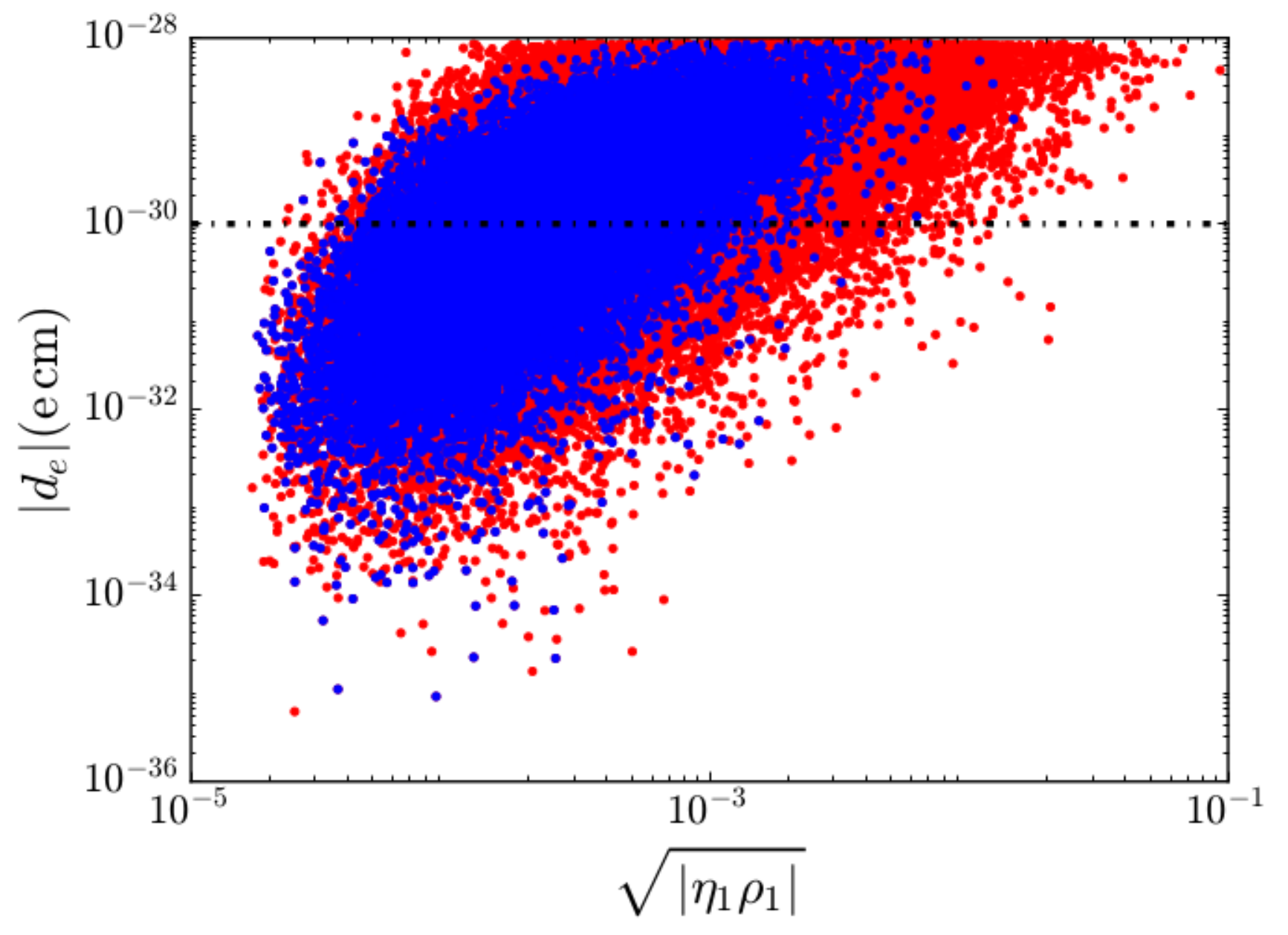}
    \caption{Expected values for the eEDM as a function of $\sqrt{|\eta_1\rho_1|}$ for NH (left panel) and IH (right panel).
      The red points constitute the current viable parameter space while the blue points are beyond the reach of future CLFV searches.}
    \label{fig:EDM1}
  \end{center}
\end{figure}

The results for the eEDM as a function of the Yukawa coupling product $\sqrt{|\rho_1\eta_1|}$ are displayed in Fig.~\ref{fig:EDM1}.
Our results show that eEDM future searches \cite{Baron:2013eja} (dashed lines) may test regions beyond the reach of experimental sensitivity of CLFV searches.
%

\section{Conclusions}
\label{sec:conclusions}

We have explored the inert Zee model in  light of the ambitious experimental program designed to probe, via charged LFV processes and EDM signals, beyond the Standard Model scenarios. 
We determined the viable parameter space consistent with the current constraints coming from DM, neutrino oscillations, LFV processes, eEDM, electroweak precision tests and collider physics.
We have also established the most relevant experimental perspectives regarding LFV searches, where we have found that $\mu-e$ conversion in muonic experiments constitutes the most promising way in this line of research.  
Furthermore, since the Yukawa couplings that reproduce the neutrino oscillation observables are complex, which in turn provide new sources of CP violation, we have shown the regions in the parameter space where the prospects for the eEDM are within the future experimental sensitivity.
It is remarkable the impact that may have eEDM future searches since they may probe the model in regions out the reach of all the future CLFV projects.

\section*{acknowledgment}

This work has been partially supported by the Sostenibilidad program of Universidad de Antioquia UdeA, CODI-E084160104 grant and by COLCIENCIAS through the grants 111565842691 and 111577657253. 


\end{document}